\documentclass[sigconf, 10pt]{acmart}
\usepackage{booktabs} % For formal tables

\usepackage{subfigure}

\usepackage{color}
\definecolor{lightgray}{rgb}{0.95, 0.95, 0.95}
\definecolor{darkgray}{rgb}{0.4, 0.4, 0.4}
\definecolor{editorGray}{rgb}{0.95, 0.95, 0.95}
\definecolor{editorOcher}{rgb}{1, 0.5, 0} % #FF7F00 -> rgb(239, 169, 0)
\definecolor{editorGreen}{rgb}{0, 0.5, 0} % #007C00 -> rgb(0, 124, 0)
\definecolor{orange}{rgb}{1,0.45,0.13}      
\definecolor{olive}{rgb}{0.17,0.59,0.20}
\definecolor{brown}{rgb}{0.69,0.31,0.31}
\definecolor{purple}{rgb}{0.38,0.18,0.81}
\definecolor{lightblue}{rgb}{0.1,0.57,0.7}
\definecolor{lightred}{rgb}{1,0.4,0.5}
\usepackage{upquote}
\usepackage{listings}
\lstdefinelanguage{JavaScript}{
  morekeywords={typeof, new, true, false, catch, function, return, null, catch, switch, var, if, in, while, do, else, case, break},
  morecomment=[s]{/*}{*/},
  morecomment=[l]//,
  morestring=[b]",
  morestring=[b]'
}
\lstdefinestyle{js} {%
  basicstyle={\footnotesize\ttfamily},   
  frame=b,
  xleftmargin={0.75cm},
  numbers=left,
  stepnumber=1,
  firstnumber=1,
  numberfirstline=true, 
  identifierstyle=\color{black},
  keywordstyle=\color{blue}\bfseries,
  ndkeywordstyle=\color{editorGreen}\bfseries,
  stringstyle=\color{editorOcher}\ttfamily,
  commentstyle=\color{brown}\ttfamily,
  language=JavaScript,
  alsodigit={.:;},  
  tabsize=3,
  showtabs=false,
  showspaces=false,
  showstringspaces=false,
  extendedchars=true,
  breaklines=true,
  literate=%
}

\setcopyright{rightsretained}
\acmDOI{10.475/123_4}

\acmISBN{123-4567-24-567/08/06}

\acmConference[SPLASH LIVE, Boston, Mass. U.S.A.]{}{November 2018}{}
\acmYear{2018}
\copyrightyear{2018}

\acmArticle{4}
\acmPrice{15.00}

\begin{document}
\title{Chalktalk : A Visualization and Communication Language}
\subtitle{As a Tool in the Domain of Computer Science Education}
%\titlenote{Produces the permission block, and
%  copyright information}
%\subtitle{Extended Abstract}
%\subtitlenote{The full version of the author's guide is available as
%  \texttt{acmart.pdf} document}

\author{Ken Perlin}
%\authornote{anonymous}
\affiliation{%
  \institution{New York University}
  \streetaddress{anonymous}
  %\city{anonymous}
  %\state{anonymous}
  \postcode{anonymous}
}
\email{ken.perlin@gmail.com}

\author{Zhenyi He}
%\authornote{The secretary disavows any knowledge of this author's actions.}
\affiliation{%
  \institution{New York University}
  \streetaddress{anonymous}
  %\city{anonymous}
  %\state{anonymous}
  \postcode{anonymous}
}
\email{zh719@nyu.edu}

\author{Karl Rosenberg}
%\authornote{This author is the one who did all the really hard work.}
\affiliation{%
  \institution{New York University}
  \streetaddress{anonymous}
  %\city{anonymous}
  %\state{anonymous}
  \postcode{anonymous}
}
\email{ktr254@nyu.edu}


\begin{abstract}
In the context of a classroom lesson, concepts must be visualized and organized in many ways depending on the needs of the teacher and students. Traditional presentation media such as the blackboard or electronic whiteboard allow for static hand-drawn images, and slideshow software may be used to generate linear sequences of text and pre-animated images. However, none of these media support the creation of dynamic visualizations that can be manipulated, combined, or re-animated in real-time, and so demonstrating new concepts or adapting to changes in the requirements of a presentation is a challenge. Thus, we propose Chalktalk as a solution. Chalktalk is an open-source presentation and visualization tool in which the user's drawings are recognized as animated and interactive "sketches," which the user controls via mouse gestures. Sketches help users demonstrate and experiment with complex ideas (e.g. computer graphics, procedural animation, logic) during a live presentation without needing to create and structure all content ahead of time. Because sketches can interoperate and be programmed to represent underlying data in multiple ways, Chalktalk presents the opportunity to visualize key concepts in computer science: especially data structures, whose data and form change over time due to the variety of interactions within a computer system. To show Chalktalk's capabilities, we have prototyped sketch implementations for binary search tree (BST) and stack (LIFO) data structures, which take advantage of sketches' ability to interact and change at run-time. We discuss these prototypes and conclude with considerations for future research using the Chalktalk platform.
\end{abstract}



\maketitle

\section{Introduction}
Especially in the context of a classroom, presenters must supplement verbal communication with visuals to illustrate concepts -- specifically those whose behavior and representations change variably or over time. Such visualizations are crucial for effective conveyance of ideas in areas such as physics, computer science, and animation, wherein ideas revolve around dynamic, interactive entities. However, traditional media such as blackboards, slide shows, and (more recently) electronic smart boards allow only for static drawings and text/image sequence, or in the best cases, fixed animations made in advance \cite{nunes2017atypical, perlin2018chalktalk}. In addition, visualizations created with these media can neither adapt to changes in the flow of the lesson nor be modified to express more complex, interrelated concepts. To address the need for a dynamic and interactive presentation system, we introduce Chalktalk.

Chalktalk is an open-source presentation and communication tool in which the user creates and manipulates interactive, animatable objects -- called "sketches" -- in real-time to demonstrate ideas\cite{perlin2018chalktalk}. Chalktalk contains a growing library of programmable sketches based on concepts from areas such as physics, mathematics, audio, computer graphics, procedural animation, and others. These can be controlled via mouse gestures and linked to form increasingly complex systems and to adapt to the changing flow of a presentation -- particularly classroom lessons. Chalktalk presents the opportunity to improve our visualization of key concepts in computer science: especially data structures, whose data and form change over time due to the variety of interactions within a computer system. Like the popular Scratch\cite{ScratchEdu} language, Chalktalk can help new programmers focus on higher level logic and concepts rather than on the syntactic peculiarities of a particular language. In addition, Chalktalk's links can transmit any data (including function callbacks) between Sketches, allowing for a more flexible system than one built for a specific domain (e.g. Max/MSP's audio modules \cite{nunes2017atypical}). These characteristics, combined with the ability to interact and build with sketches in real time, make Chalktalk a promising environment in which to learn and explore computer science. Thus, we propose a computer science-centric Chalktalk library and contribute prototype sketches based on fundamental data structures -- the binary search tree (BST) and stack. We will use these prototypes as the basis for further investigation of alternative visualizations and interactions for use in computer science education. Here we provide an overview of Chalktalk's system and use cases, then discuss our prototype data structure sketch implementations in detail. We conclude with notes on ongoing and future research using the Chalktalk platform.

\section{Presenting with Chalktalk}
\subsection{Example Use Case}

\begin{figure}[h]
\includegraphics[width=0.45\textwidth]{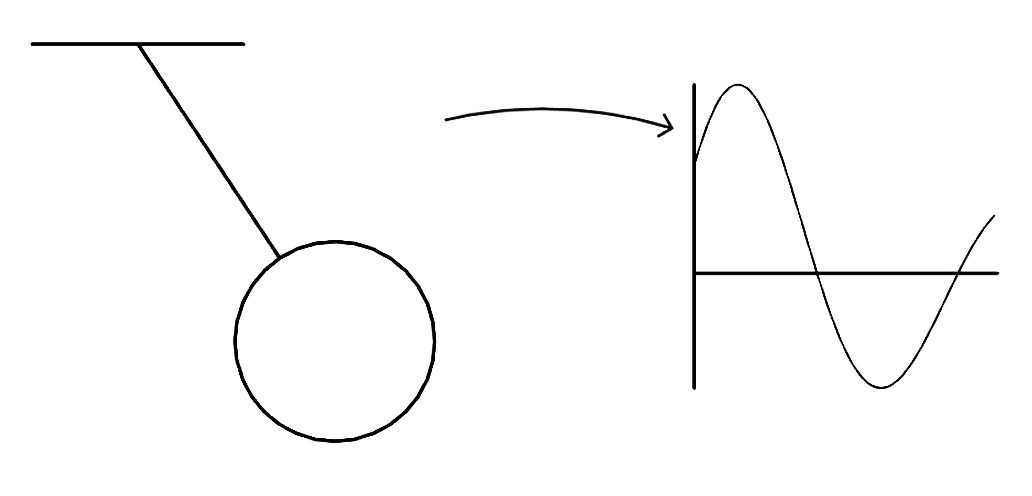}
\caption{The user has dragged the mouse to swing the pendulum (left), which outputs its angle as numerical data for the graph (right) to display as a curve. In this case, the curve represents the pendulum's displacement. However, the graph can interpret all numerical data it receives, which means any sketch that outputs numbers can interact with the graph sketch.}
\label{fig:1a}
\end{figure}

\begin{figure}[t!]
\includegraphics[width=0.45\textwidth]{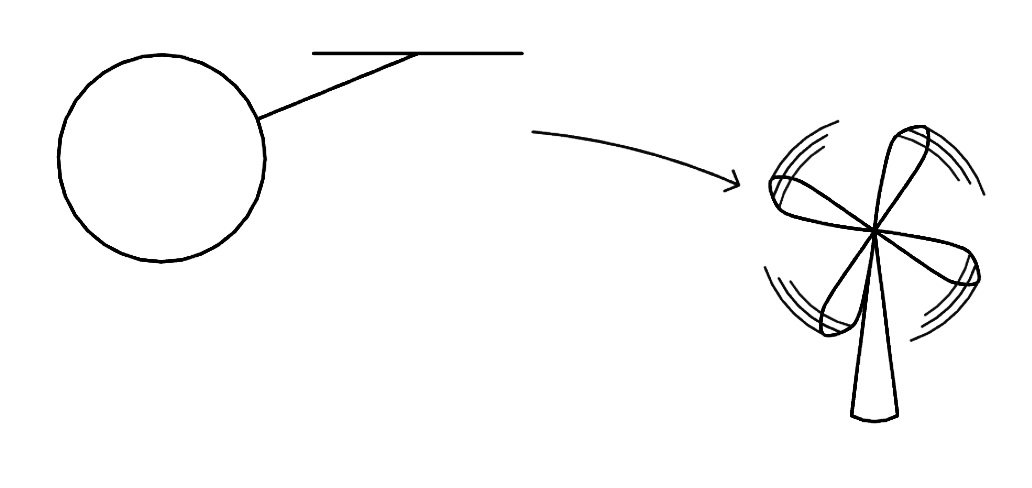}
\caption{The same pendulum is linked to a fan sketch (right), which uses the data from the pendulum to set its own angle, causing it to rotate. Neither sketch is aware of each other's types. Each merely sends and receives data that are interpreted independently.}
\label{fig:1b}
\end{figure}

\begin{figure}[hb]
\includegraphics[width=0.45\textwidth]{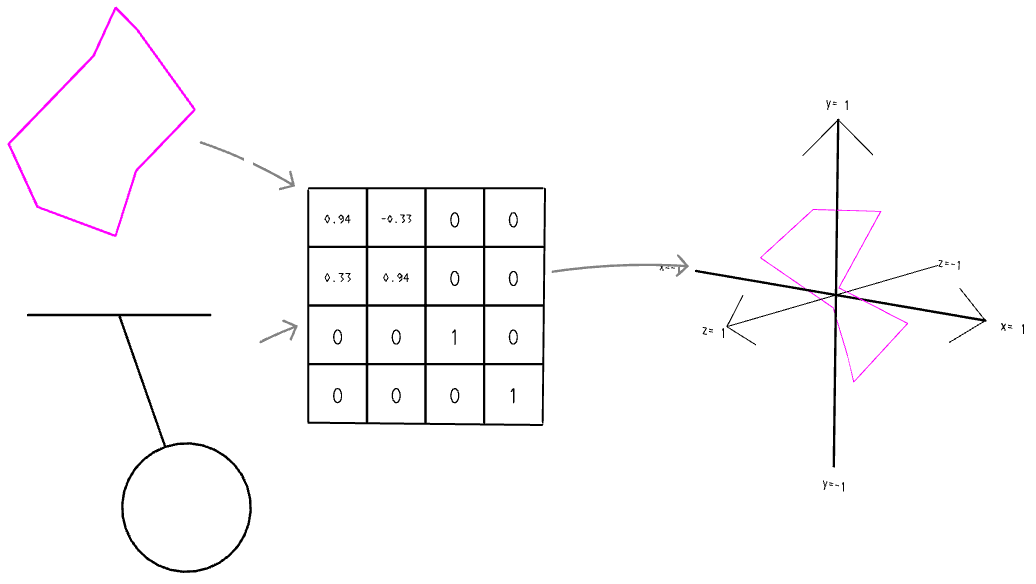}
\caption{In this case, the matrix and coordinate system, too, process only data transferred between sketches -- not the sketches themselves. This data-oriented design allows for the flexible combination of and interaction between sketches.
Top-left: a 3D shape whose output is mesh data; 
Bottom-left: another pendulum; middle: a transformation matrix set to rotate on z; 
Right: a 3D coordinate system -- The matrix receives an angle from the swinging pendulum, which it uses to set the rotation. The matrix processes, transforms, and outputs the shape data, which the coordinate system displays. }
\label{fig:1c}
\end{figure}

A simple example use case\cite{nunes2017atypical} is as follows:

Suppose a physics teacher wishes to illustrate the motion of a pendulum. The teacher should be able to draw the pendulum and swing it to demonstrate its physical properties. In addition, the teacher should be able to draw a graph to which she can link the pendulum to show the mathematical curve representing its displacement from equilibrium (see figure~\ref{fig:1a}). Furthermore, this pendulum sketch should be reusable in a network of other sketches to allow for further experimentation -- perhaps as a controller for the movement of other objects such as a fan (see figure~\ref{fig:1b}) or for the configuration of a matrix to rotate 3D geometry (see figure~\ref{fig:1c}).

\subsection{USER AND PROGRAMMER INTERFACE}
\subsubsection{RECOGNITION}
\begin{figure}[h]
\includegraphics[width=0.45\textwidth]{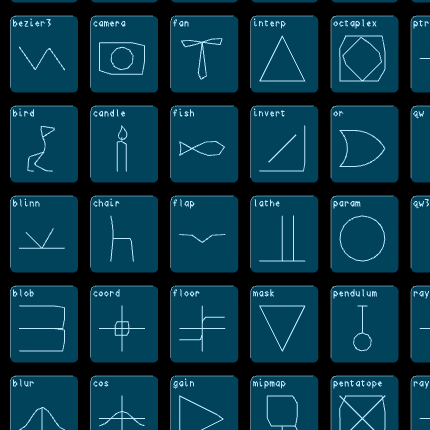}
\caption{A cross-section of Chalktalk's library, containing mathematical functions, creatures, geometry, and other entities and concepts}
\label{fig:1d}
\end{figure}

To instantiate a sketch, the user first free-hand draws a glyph, composed of a specific number and ordering of strokes. That glyph is compared against a library of glyphs (see figure~\ref{fig:1d}) defined in Chalktalk, and the closest match is selected for recognition. Finally, the user clicks the recognized glyph to instantiate it as a sketch.

\subsubsection{USER INPUT}
Sketches are able to recognize swipe motions, clicks, and drags as input events. Callback functions defined in a sketch can be used to  trigger events in response to mouse gestures. There is also a general set of "command" gestures recognized by all sketches, which can be used to modify the scale, position, rotation, and other properties of the sketch itself. These command gestures always begin with a click around the periphery of the sketch.

\subsubsection{LINKS AND DATA TRANSFER}
Links transfer data from one sketch to another. Connecting a link from and to a sketch is a matter of wiring the link visually using a "command drag." This is achieved by clicking on the source sketch's left periphery, and then clicking and dragging an arrow from the source to the destination sketch. In the internal JavaScript code, an output procedure returns data for receipt by any number of other linked sketches, and each sketch may also directly access data sent to it.

\subsubsection{SKETCH DESIGN AND IMPLEMENTATION}
\begin{figure}[h]
\centering
\includegraphics[width=0.45\textwidth]{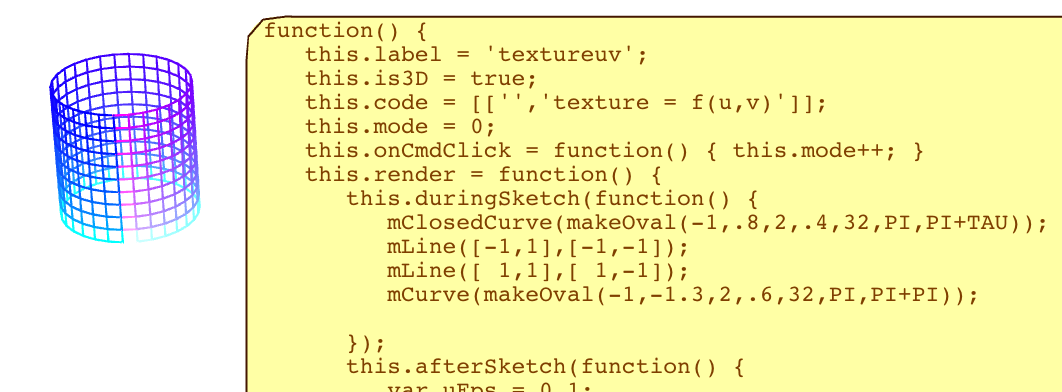}
\caption{All sketch codes can be exposed in the Chalktalk interface. On the left is a 3D cylindrical wire-frame mesh. On the right is a fragment of its code, which can be edited live.}
\label{fig:2d}
\end{figure}

Because Chalktalk is open-source, users may choose to use existing sketches or to design their own, either for personal use or to contribute to the growing repository. This means that users (e.g. teachers and students) need not be programmers. Nevertheless, one of Chalktalk's strengths lies in its programmability. Chalktalk runs in the browser, so sketches are written as JavaScript files in which swipe, drag, output, render loop, and other methods attached to that sketch type are defined. Designing the appearance of a sketch as well as its glyph (used for recognition) involves calling draw functions that specify curves, colors, and other attributes or applying matrix transformations. These functions take the form of mCurve(..), mLine(..), mOval(..), m.translate(..), m.scale(..), color(..) and others. The drawing API may be comparable to Processing's\cite{ProcessingHandbook}, as it allows the programmer to think in terms of curves and shapes rather than low-level draw calls. As a user becomes more proficient in programming practices, all JavaScript language constructs (conditionals, loops, variables) may be used to adjust the appearance and behavior of a sketch. Methods such as this.output() may be defined in a sketch to send data across links, and within the Chalktalk interface, the user may also access a sketch instance's code for viewing or editing at run-time (see figure~\ref{fig:2d}).

\subsection{Narrative and Performance}

Chalktalk is also a performative medium in which "programming" is not only the literal programming of sketches, but also the narrative that unfolds when combining and juxtaposing sketches in real-time. For example, a Chalktalk-based computer science library would include sketches that might simulate pieces of a program. A teacher could begin the conversation with her students by using sketches to introduce fundamental programming constructs such as the loop. Based on student engagement, she could then arrange, link, and compose the sketches into increasingly specialized systems of data structures and logic, mirroring the way in which traditional lessons build on previous concepts. The teacher may also invite students to design their own experiments based on the day's lesson. This sense of engagement and conversation is key.

\section{prototype: computer science data structure sketches}

We believe Chalktalk to be an ecosystem particularly fitting for the visualization of computer science concepts such as data structures, which may take on multiple representations and behaviors depending on use and composition. To show the potential of a computer science sketch library, we present two work-in-progress data structures sketches -- a binary search tree (BST) and stack -- that illustrate the interoperability and interactivity of sketches.

\subsection{Binary Search Tree}

\begin{figure}[h]
\centering
\subfigure[The glyph that is recognized as a BST sketch]{\includegraphics[width=0.23\textwidth]{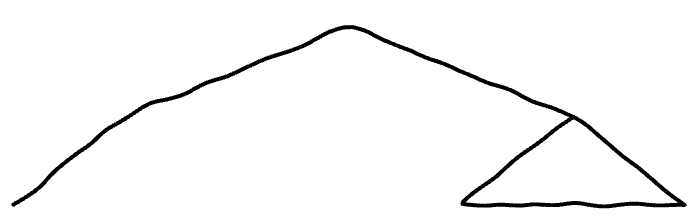}\label{fig:6a}}
\subfigure[The BST sketch in its initial state, pre-populated with nodes]{\includegraphics[width=0.23\textwidth]{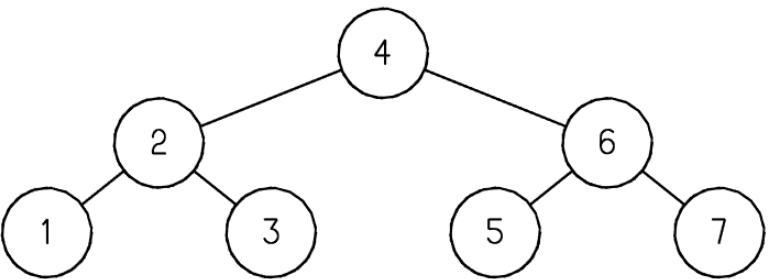}\label{fig:6b}}
\caption{Binary Search Tree Sketch Recognition}\label{fig:6}
\end{figure}

% \begin{figure}[h]
% \includegraphics[width=0.5\textwidth]{small_tree_w}
% \caption{The binary search tree in its initial state, pre-populated with nodes}
% \label{fig:3a}
% \end{figure}
\begin{figure}[b!]
\includegraphics[width=0.4\textwidth]{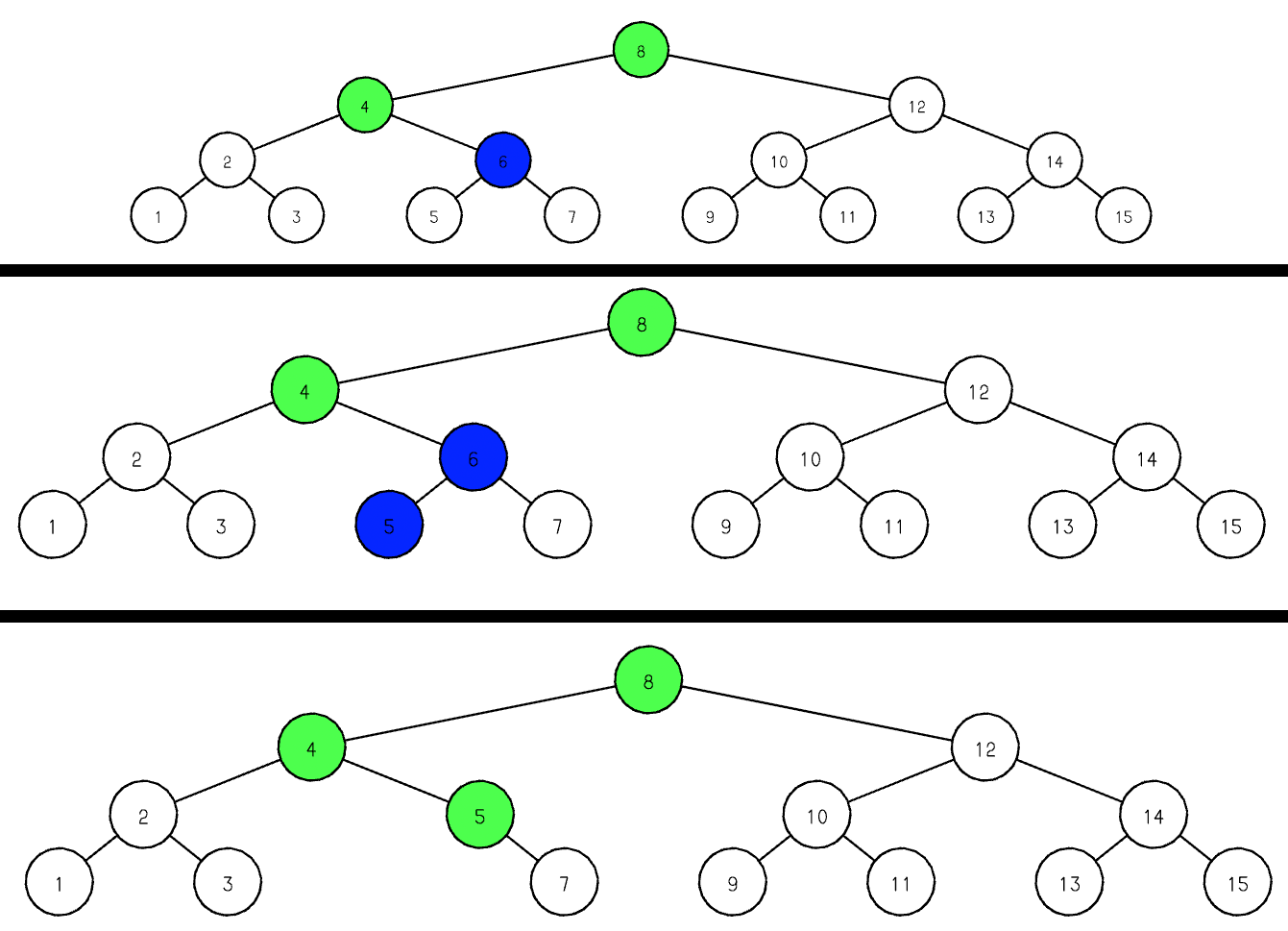}
\caption{The node removal visualization for the BST sketch -- top: because the user has selected the now-blue node for removal, the algorithm has recursed down to that node; middle: since the node has two children, the algorithm searches for the predecessor node for a replacement; bottom: the blue node has been replaced with its predecessor (note that this recursion sequence is animated)}
\label{fig:3ai}
\end{figure}

The internal code of the BST sketch (figure ~\ref{fig:6}) contains a reference-based tree implementation that the user modifies via mouse interactions with the sketch, which map to operations such as insert, remove, and preorder-, in-order-, post-order-, and breadth-first traversal. Operations may also be undone using a leftward swipe. Insertion and removal are initiated by creating a numerical sketch (built into Chalktalk) and dragging and dropping it onto the tree. This starts the recursive process of searching for the proper node, whereby the in-memory tree performs the algorithm while issuing draw commands to highlight visited nodes in the Chalktalk interface. Removal is implemented in a similar manner, but accounts for the additional cases in the removal algorithm by providing additional animations. For example, when the node to remove has two child nodes, the algorithm must find the predecessor or successor node to substitute (see figure ~\ref{fig:3ai}).

\begin{figure}[h]
\includegraphics[width=0.4\textwidth]{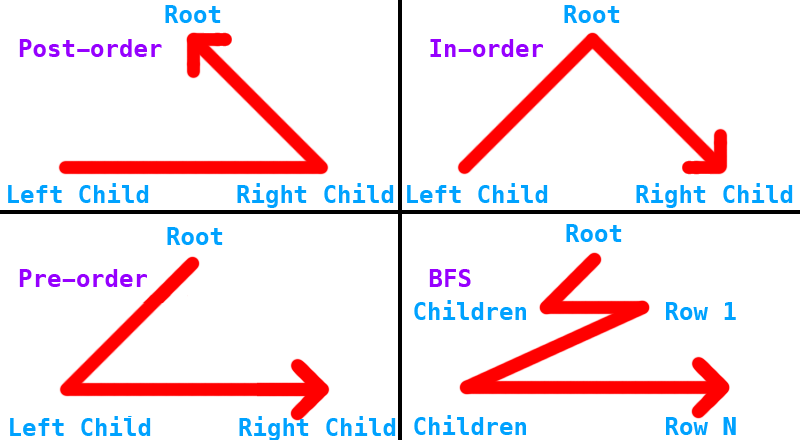}
\caption{Seen here are the four BST-specific gestures that map to the different traversals the sketch supports. They are meant to serve as (experimental) visual mnemonic devices for learning purposes. For example, in post-order traversal, all children are visited first, so this is represented as an arrow moving from left child to right child, and then to the root. (For in-order traversal, the arrow moves from the left child to the root node, then to the right child. In Pre-order traversal the arrow moves from the root to the left child, and then to the right child.) Breadth-first search (BFS) traverses the tree in layers, so a zig zag from root to child represents this process. We are interested in pursuing research related to these sorts of mnemonic devices.}
\label{fig:3b}
\end{figure}

\begin{figure}[h]
\includegraphics[width=0.4\textwidth]{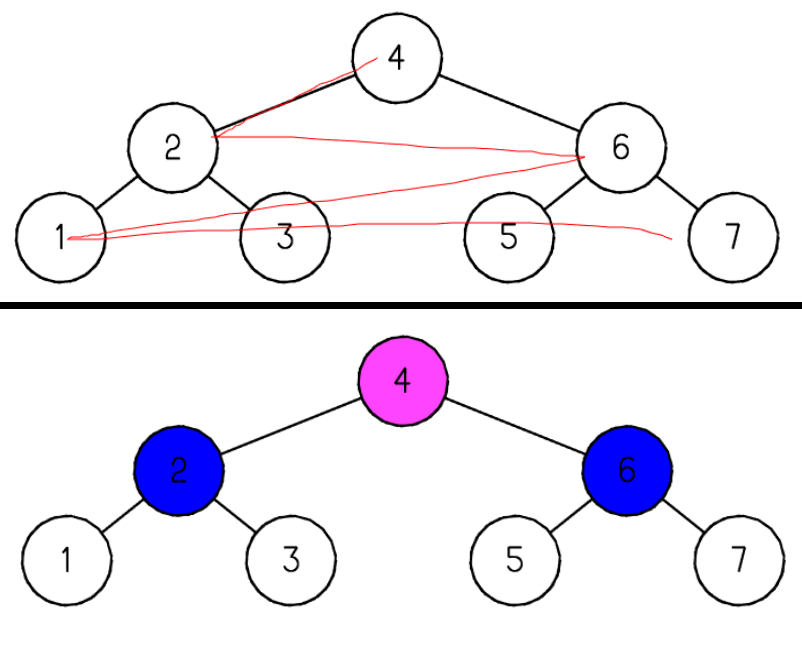}
\caption{top: the user has drawn a zig-zag curve atop the BST to start a breadth-first search traversal; bottom: the traversal is underway -- the algorithm is now visiting the first row of child nodes (in blue)}
\label{fig:3c}
\end{figure}

When demonstrating a tree traversal, one way of appealing to visual learners might be to represent the sequence of parent/child node visits with a mouse gesture. The presenter would draw a special curve (see figure~\ref{fig:3b} for the design and figure~\ref{fig:3c} for an example) to trigger its corresponding traversal. This idea is implemented into the BST sketch by reusing the functionality of Chalktalk's sketch recognition system for the recognition of line strokes drawn atop the BST.

All interactive operations require timed pauses and interpolated movements across multiple frames, so functions responsible for modifying the tree make use of JavaScript's "yield" keyword and generators to save state when an operation is in progress (e.g. tree traversals). To supplement the raw language features and provide higher level interface, a small host of utility functions for timing and pausing was written.
To run and animate an operation, the BST sketch enqueues the generator function specific to a given operation, runs the generator until a pause in the animation is necessary (a yield), renders the updated sketch, and checks the queue on the next frame to resume execution of the function. This process repeats until the current operation has been completed. Objects modeled after debug breakpoints can be inserted into the code path of a generator to allow the user to pause deliberately.

\subsection{Stack}
\begin{figure}[h]
\includegraphics[width=0.4\textwidth]{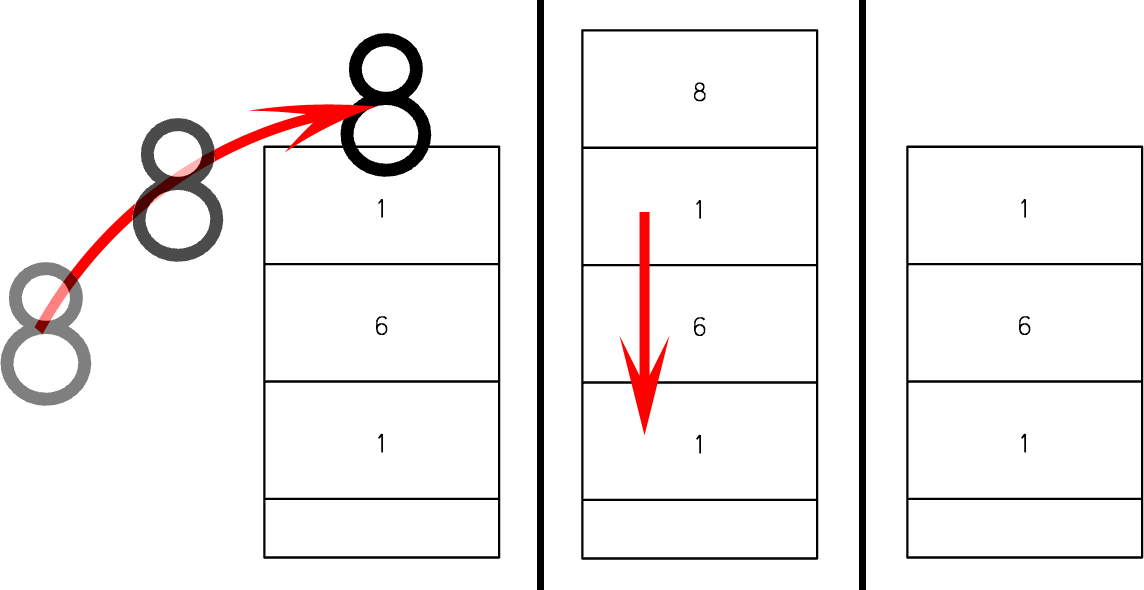}
\caption{left: the user drags and drops an "8" sketch onto a stack sketch containing the values 1, 6, 1; middle: the 8 has been pushed to the top of the stack, the user now clicks and drags downwards to pop the stack; right: the stack has been returned to its previous state}
\label{fig:3d}
\end{figure}

The LIFO stack sketch uses a JavaScript array internally as well as generators to save intermediary state over multiple frames, just as the BST does. To push a value, the user draws, for example, a numerical sketch, and drags-and-drops it atop the stack. To pop, the user does a downward swipe gesture with the mouse (see figure~\ref{fig:3d}).

% \begin{figure}[h]
% \centering
% \subfigure[the user drags and drops an "8" sketch onto a stack containing the values 1, 6, 1]{
% 	\includegraphics[height=0.2\textwidth]{stack_a}\label{fig:10a}}\quad
% \subfigure[the 8 has been pushed to the top of the stack, the user now clicks and drags downwards to pop the stack]{
% 	\includegraphics[height=0.2\textwidth]{stack_b}\label{fig:10b}}\quad
% \subfigure[the stack has been returned to its previous state]{
% 	\includegraphics[height=0.2\textwidth]{stack_c}\label{fig:10c}}\\
% \caption{???}
% \end{figure}

If a sketch is linked to the stack, additional logic will be used to decide whether it is necessary to push or pop, as the link might send the same data repeatedly. Duplicate pushes and pops of that same data should be ignored.

\subsection{Final Expected Behavior}

\begin{figure}[hb]
\includegraphics[width=0.4\textwidth]{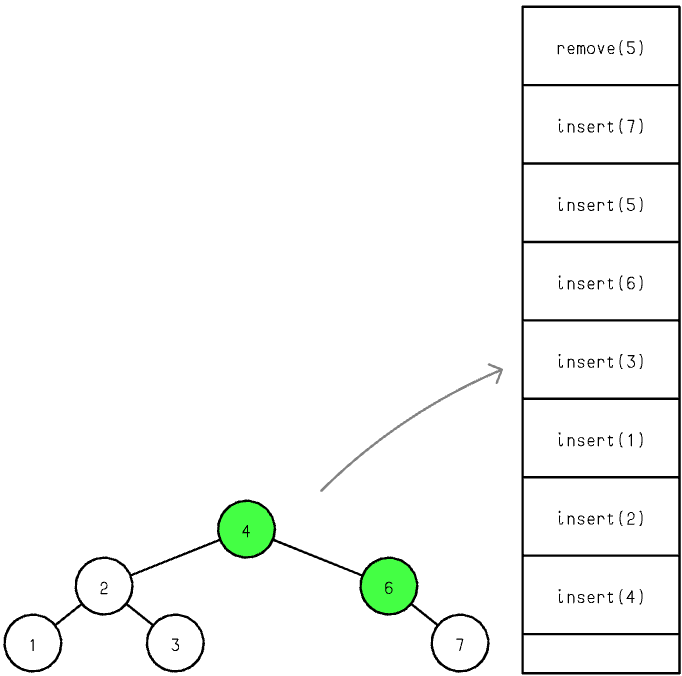}
\caption{The tree currently outputs a record of the operations it performs. Because the stack can accept any form of numerical or string data, it can display these records as a history. It can also interact with any sketch in the Chalktalk world that sends compatible data. At the top of the stack seen here, a remove(5) record indicates that the node with value 5 has just been removed. When complete, the BST will output information that the stack can use to simulate a call stack during a recursive algorithm.}
\label{fig:3e}
\end{figure}

The final, completed versions of the BST and stack sketches will exhibit new behavior when linked: during a traversal, the BST will output information about the most recently visited or exited node, and the stack will interpret this data as stack frames to push and pop. As a result, the linked tree and stack will show the relationship between recursion and a call-stack. The current in-progress implementations interact differently for now: the tree outputs a record of the most recently performed operation insertion or removal, and the stack displays the history of these records (see figure~\ref{fig:3e}).

Thus, the BST and stack have new behaviors when linked, just like the pendulum and graph. Once implemented, arrays, FIFO queues, graphs, and other data structures could could be used to interact in a larger high level simulation of a program or with other compatible sketches such as matrices and vectors.

\section{Ongoing and Future Research}

One of the main strengths of Chalktalk's system is that it gives the user the freedom to build and use sketches in a variety of ways. In-development Chalktalk features such as improved code hot-loading and property editing will likely impact the way in which sketches are designed, and as a result, will lead to changes in the proposed data structures library as we experiment with different implementations.
Chalktalk has, in fact, been evolving. It lies at the center of research that includes the development of a type system for Chalktalk's links\cite{nunes2017atypical} and the creation of interactive augmented reality (AR) and virtual reality (VR) environments \cite{perlin2018chalktalk}. In addition, Chalktalk is being used in classrooms to teach computer graphics, animation, sound processing and other subjects \cite{perlin2016future}. In the near future, we also wish to conduct case studies on a larger scale.

\section{Conclusion}

We have introduced Chalktalk, a presentation and visualization tool for learning, exploration, and communication. By prototyping data structure sketches we have shown how a computer science sketch library might be implemented to facilitate dynamic lessons and discussions. Chalktalk, we believe, has great potential as an educational tool, especially in the domain of programming and computer science.

%\end{document}  % This is where a 'short' article might terminate

\bibliographystyle{ACM-Reference-Format}
\bibliography{sample-bibliography}

\end{document}